\documentclass[letterpaper, 10 pt, conference]{ieeeconf}  % Comment this line out if you need a4paper

\usepackage{fullpage}
\usepackage{graphicx,latexsym}
\usepackage{epsfig}
\usepackage{amsmath,amssymb}

\newcommand{\done}{\hspace*{\fill} $\Box$}
\newcommand{\RE}{{\mathbb R}}
\newtheorem{theorem}{Theorem}[section]

\newtheorem{corollary}[theorem]{Corollary}
\newtheorem{lemma}[theorem]{Lemma}
\newtheorem{remark}[theorem]{Remark}
\newtheorem{definition}[theorem]{Definition}

\newtheorem{example}{Example}[section]

\title{\bf From LQR to Static Output Feedback: a New LMI Approach}
\author{
Luis Rodrigues
\\
\small Department of Electrical and Computer Engineering \\
\small Concordia University \\
\small Montr\'eal, QC, Canada\\
\small Email: {\tt luis.rodrigues@concordia.ca}
}

\begin{document}
\pagestyle{empty}
\maketitle

\begin{abstract}
This paper proposes a new Linear Matrix Inequality
(LMI) for static output feedback control assuming that a Linear
Quadratic Regulator (LQR) has been previously designed for
the system. The main idea is to use a quadratic candidate
Lyapunov function for the closed-loop system parameterized
by the unique positive definite matrix that solves the Riccati
equation. A converse result will also be proved guaranteeing
the existence of matrices satisfying the LMI if the system
is static output feedback stabilizable. 
\iffalse
The proposed method
will then be extended to the design of static output feedback
for the $H_{\infty}$ control problem. Besides being a sufficient
condition for which a converse result is proved, 
There are four main advantages of the proposed methodology. 
First, it is computationally tractable.
Second, one can use weighting matrices and obtain a solution
in a similar way to LQR design.
Third, the proposed method has an extremely simple LMI structure 
when compared with other LMI methods proposed in the literature.
Finally, for the cases where the output is equal to the
state it is shown that the LQR solution verifies the proposed
LMI. Therefore,
\fi
The static output feedback includes the LQR
solution as a special case when the state is available, which is
a desirable property. The examples show that the method is
successful and works well in practice.
\end{abstract}

\section{Introduction}
Static output feedback for linear systems is still
an open problem in the sense that no satisfactory computationally
tractable necessary and sufficient condition has been
found for a system of arbitrary order. Although some specialized
cases of single input single output systems of lower
order have been solved successfully with tractable methods
(see \cite{astolfi00} and references therein), static output feedback is still
an important research topic in control theory that
dates back to the 1960s \cite{davison69}. 
\iffalse
The pioneering research efforts
of the 1960s did not have however the powerful tool of LMIs
at their disposal. 
It appears that Willems in 1971, who coined
the term LMI, was also the first to recognize that LMIs had
the potential to be solved by computational algorithms when
he wrote \cite{willems71}: "The basic importance of the LMI seems to be
largely unappreciated. It would be interesting to see whether
or not it can be exploited in computational algorithms, for
example." 
It was not until the end of the 1980s that interior
point algorithms were first proposed in the literature \cite{nesterov94},
which were later used to solve LMIs \cite{boyd94} numerically in
software packages~\cite{cvx}.
\fi

The problem of static output feedback re-emerged in the 1990s with a seminal paper by Trofino and Ku\u{c}era \cite{trofino93}.
Shortly after, in a survey on the state of systems and control \cite{blondel95} it was stated that finding a tractable efficient algorithm for static output feedback stabilization was an open problem of control theory.
Another survey on static output feedback~\cite{grigoriadis97} 
stated that the static output feedback
multivariable problem was "still analytically unsolved" and
that "no efficient algorithmically solutions" were available.
\iffalse
That survey also shows that the problem of dynamic output
feedback can be brought to a static output feedback formulation
in an augmented state space. Therefore, besides
having an interest on its own, the problem of static output
feedback also sheds light into dynamic output feedback.
\fi
Since the surveys \cite{blondel95,grigoriadis97} were published many researchers
have proposed different algorithms for static output feedback
stabilization. However, up to the end of the 1990s most
of the methodologies proposed in the literature were nonconvex and involved
iterative procedures. An iterative algorithm with guaranteed
linear quadratic suboptimality was proposed in \cite{geromel94}. 
A cone complementarity linearization algorithm was presented in reference
\cite{elghaoui97} that was applied with promising results to a
collection of numerical examples with randomly generated
parameters. In reference \cite{cao98} the authors propose an iterative
LMI method to compute the feedback gain for static output
feedback controllers. The iterative algorithm published in
\cite{veseley01} uses the same structure for the gain matrix proposed
in \cite{geromel85} yielding a static output feedback interpretation as
an equivalent state feedback. 

It appears that the first time that a semidefinite programming
solution was proposed for the related problem of the least order dynamic
output feedback synthesis was in reference \cite{mesbahi99}.
In that paper the elimination lemma and matrix dilations were used to reduce the solution of a Bilinear Matrix Inequality (BMI) to a convex semi-definite program. 
However, although the proposed algorithm is not iterative, it is nonetheless 
quite intricate involving seven different steps.
One year later, reference \cite{astolfi00} addressed static output feedback stabilization for both
linear and nonlinear systems. It offered a convex formulation
of necessary and sufficient conditions of stabilizability under
certain constraints on the system matrices for linear systems
of order less or equal to three.
\iffalse
, provided that the state model
is in observability canonical form.
\fi
Connections of static output feedback to LQR control were also studied in the literature.
An early contribution making a connection of static output feedback
to the solution of Ricatti equations and the
LQR problem can be found in reference \cite{geromel85}. 
Unfortunately, the results of that paper do not seem to allow a convex formulation of
static output feeback control in the form of an LMI.
The authors have therefore proposed an iterative algorithm
for synthesizing the controllers.
An interesting summary of other work connecting static output feedback with LQR up
to 2003 is included in reference \cite{mok03}. In that reference an
approach relying on the existence of two weighting matrices
fulfilling an eigenvalue inequality is proposed. 
A matrix inequality is also proposed but it depends on the product
of matrix variables in a form that does not seem to be
transformable into an LMI. In the last decade the topic
of LMIs for static output feedback synthesis regained interest in the literature.
For the most important contributions up to 2015 see the survey \cite{sadabadi16} 
or \cite{bacciotti17} for a summary of past contributions.
\iffalse
In \cite{karimi13} a two-step LMI procedure is proposed to design a
static output feedback controller which is applied to the $H_{\infty}$
control problem. However, the feedback gain is not obtained 
directly as the solution of the LMIs
because the proposed method involves a transformation of
variables for which the invese must be computed to obtain
the feedback gain after the LMI is solved. 
More recently, reference \cite{madeira16} presents a
two-step procedure involving an LMI and one analytical
formula to design a static output feedback controller
based on passivity indices and applies it successfully to two
examples.

To the best of the author's knowledge there is still no
tractable algorithm to design static output feedback controllers
for multivariable linear systems based on linear
quadratic optimal control. Motivated by this gap in the
literature this paper presents a new LMI for static output
feedback stabilization that is inspired by linear quadratic
optimal control. The derivation of the new LMI is based
on the assumption that a state feedback controller has
been previously designed by solving a Riccati algebraic
equation corresponding to an LQR problem. A converse
result will also be proved guaranteeing the existence of
matrices verifying the LMI if the system is static output
feedback stabilizable. 
\fi

The paper is organized as follows. Section \ref{mainresult} states
and proves a result on the design of stabilizing static output feedback controllers. 
The proposed methodology is then extended to systems with direct feedthrough in section \ref{directfeedthrough}.
\iffalse
 $H_{\infty}$ static output feedback 
control in section \ref{hinfinity}. 
\fi
Examples are presented in section \ref{examples}.
\iffalse
 to show that
the method is consistently successful and works extremely well in practice.
The paper uses standard notation for matrix inequalities where $I_n$ denotes the $n\times n$ identity matrix and $A>B,$ with $A=A^T, B=B^T,$ indicates that all eigenvalues of $A-B$ are positive.
\fi
\section{Static Output Feedback}\label{mainresult}
\noindent Consider the dynamics of a linear time-invariant system with measurements contaminated by a disturbance,
\begin{eqnarray}\label{linear}
\dot x &=& Ax + Bu\nonumber\\
y &=& Cx+Dw
\end{eqnarray}
where $x\in\RE^n$ is the state, $u\in\RE^m$ is the input, $w\in\RE^{n_w}$ is the disturbance, and $y\in\RE^p$ is the output.
The pair $(A,B)$ is assumed to be controllable and the pair $(A,C)$ is assumed to be observable.
The static output feedback problem in this paper is to find a control input
\begin{equation}\label{sofinput}
u=Fy=FCx+FDw
\end{equation}
such that the closed-loop system
\begin{equation}
\dot x = (A+BFC)x+BFDw
\end{equation}
is dissipative, which implies exponential stability of the closed-loop linear system in the absence of a disturbance.
Roughly speaking, a system is considered dissipative if the amount of energy that the system can provide to its environment is less than what it receives from external sources as defined next.
\vspace{5pt}

\begin{definition}\label{dissipative}
The system (\ref{linear}) is \emph{dissipative} with supply rate $W(w)$ and class $C^1$ storage function $V(x)$, if $V(x)$ is positive definite and if
\begin{equation}\label{dissp}
\dot V(x(t)) < W(w(t))
\end{equation}
When both $V$ and $W$ are quadratic functions the system is said to be quadratically dissipative.\done
\end{definition}  
\vspace{5pt}

\noindent The main idea of this paper is to design a static output feedback controller that can be obtained from a state feedback controller $u=Kx$ that has been previously designed by LQR.
This is a similar idea as the one used when designing an LQG dynamic output feedback invoking the {\em certainty equivalence} principle.
For the LQG design the feedback gain can be obtained first by LQR and then an observer is designed independently to estimate the state in accordance with the {\em separation principle}.
For the case of static output feedback no observer is needed.
For linear systems the LQR is a state feedback controller that minimizes
\begin{equation}\label{LQRfunctional}
J=\frac{1}{2}\int_0^{\infty} \left[x(t)^TQx(t)+u(t)^TRu(t)\right]dt,
\end{equation}
where $Q=Q^T\ge 0$ is a state weighting matrix and $R=R^T>0$ is a control weighting matrix, both given by the designer.
If the pair $(A,B)$ is controllable, given matrices $Q\ge 0, R>0,$ such that the pair $(A,\sqrt{Q})$ is observable, one can compute the unique positive definite matrix $P$ that is the solution of the Riccati equation
\begin{equation}\label{Riccati}
-A^TP-PA+PBR^{-1}B^TP=Q
\end{equation}
yielding the stabilizing state feedback controller
\begin{equation}\label{LQRcontrol}
u_s(t)=Kx(t)=-R^{-1}B^TPx(t)
\end{equation}
The solution of the Riccati equation can also be computed by solving the following semi-definite program (SDP), whose LMI was suggested by \cite{willems71} and the opimization functional appeared in \cite{geromel85}.
\vspace{5pt}

\noindent {\bf LQR Feedback Optimization}
\begin{equation}\label{LMILQR}
\begin{array}{ll}
\mbox {\rm max}~{\rm trace}(P)\\
s.t.  
\left[
\begin{array}{cc}
A^TP+PA+Q & PB\\
B^TP & R
\end{array}
\right] \ge 0
\end{array}
\end{equation}
%\vspace{10pt}

\noindent It will be shown in this paper that the static output gain matrix $F$ can be obtained using a Lyapunov function parameterized by the unique positive definite matrix that solves the Riccati equation for the LQR problem.
\vspace{5pt}

\noindent {\bf Problem Statement}
Using weighting matrices $Q\ge 0$ and $R>0$ design a static output feedback controller $u=Fy$ such  that when $C=I$ and $D=0$ the LQR state feedback controller naturally becomes a solution of the static output feedback problem, i.e, $F=K$ when $C=I$ and $D=0$. The closed-loop system should be dissipative in the presence of disturbances and exponentially stable in the absence of disturbances.
\vspace{5pt}

\noindent An obvious question to ask is if the state feedback gain $K$ can have the structure of the multiplication of a matrix $F$ by the matrix $C$, i.e, if $K=FC$ is a possible solution of the LQR problem.
Since from (\ref{LQRcontrol}) we know that $K=-R^{-1}B^TP,$ after multiplying on the left by $R$ this corresponds to the constraint
\begin{equation}\label{LQRsofconstraint}
RFC+B^TP=0
\end{equation}
which is a linear constraint in both the matrix $F$ and the matrix $P$.
The following result shows that there are some cases where this constraint cannot be satisfied for $P>0$ that solves the Riccati equation (\ref{Riccati}).
\vspace{5pt}

\begin{lemma}\label{sofimpossible}
Assume that the pair $(A,B)$ is controllable, and that the matrices $Q\ge 0, R>0,$ are such that the pair $(A,\sqrt{Q})$ is observable.
If $CB=0$ then the controller gain matrix which is the solution of the LQR problem does not have the structure $K=FC$.
\end{lemma}
%\vspace{10pt}
\proof
The proof then follows by contradiction.
Assume that the statement of the theorem is not true. Then from the constraint (\ref{LQRsofconstraint}) 
one can write that $B^TPB=-RFCB$. Since by assumption $CB=0$ this means that $B^TPB=0$. Since $P$ is the unique positive definite solution of the Riccati equation (\ref{Riccati}) this is not possible.
\done
%\vspace{10pt}
\begin{remark}
Note that the product $CB$ is an invariant of the system, which is independent of the state space realization.
Therefore, one can use the state variables leading to the controllability canonical form, without loss of generality.
Doing so, one can easily prove that for SISO systems $CB=0$ corresponds to saying that the transfer function of the open-loop system has relative degree two or higher.\done
\end{remark}
%\vspace{10pt}

%\noindent The following Schur complement lemma will be used in the proofs of the main results.
\vspace{5pt}
\begin{lemma}\cite{boyd94}\label{Schurlemma}
Let matrices $X(x), Y(x), Z(x)$ be affinely dependent on $x$ with proper dimensions where $X(x)=X(x)^T,~Z(x)=Z(x)^T,$ and define
\begin{eqnarray*}
M(x)=\left[
\begin{array}{lr}
X(x) & Y(x)\\
Y^T(x) & Z(x)\\
\end{array}
\right]
\end{eqnarray*}
If $Z(x)$ is invertible then 
\begin{equation}
\mathbb{S_Z}(x)=X(x)-Y(x)Z(x)^{-1}Y^T(x)
\end{equation}
is called the Schur complement of $Z(x)$.
The following statements are true:
\begin{enumerate}
  \item $M(x)>0$ if and only if $Z(x)>0$ and $\mathbb{S_Z}(x)>0$,
  \item $M(x)\ge 0$ if and only if $Z(x)\ge 0$,
\begin{equation}
X(x)-Y(x)Z^{\dagger}(x)Y^T(x)\ge 0,
\end{equation}
and $Y(x)\left[I-Z(x)Z^{\dagger}(x)\right]=0$, where $Z^{\dagger}(x)$ denotes the Moore-Penrose inverse of $Z(x)$.
Furthermore, when $Z(x)>0$ then $M(x)\ge 0$ if and only if $\mathbb{S_Z}(x)\ge 0$.
\end{enumerate}
\end{lemma}
\vspace{5pt}

\noindent We start by considering the case $D=0$.
\vspace{5pt}

\begin{theorem}\label{softheorem0}
Assume that the linear system
\begin{eqnarray}\label{linearnoD}
\dot x &=& Ax + Bu\nonumber\\
y &=& Cx
\end{eqnarray}
 is given with $(A,B)$ controllable and $(A,C)$ observable.
Assume further that $Q=Q^T\ge 0, R=R^T>0,$ are such that the pair $(A,\sqrt{Q})$ is observable.
If there exists a matrix $F$ verifying the LMI
\begin{equation}\label{LMIforstability}
(A+BFC)^TP+P(A+BFC)<0
\end{equation}
where $P>0$ is the unique positive definite solution of the Riccati equation (\ref{Riccati}) then the control law $u=Fy$ is a stabilizing static output feedback for system (\ref{linearnoD}).
Conversely, if the system (\ref{linearnoD}) is static output feedback stabilizable then there always exit $F, P>0$ verifying LMI (\ref{LMIforstability}).
\end{theorem}
%\vspace{10pt}
\proof
To prove sufficiency note that LMI (\ref{LMIforstability}) corresponds to the the sufficient condition for stability of the closed-loop system under the feedback law $u=Fy$ using as Lyapunov function $V=x^TPx$.
To prove the converse result, assume that system (\ref{linearnoD}) is stabilizable by static output feedback.
Since the system is linear it is necessary to find a quadratic Lyapunov function for the closed-loop system, i.e, it is necessary that there exist matrices $P>0$ and $F$ verifying (\ref{LMIforstability}).
\done
\vspace{5pt}

\noindent The next result is valid for the case $D\neq 0$ and includes the weighting matrices $Q, R$ in the design of $F$. These matrices can then be used as tuning parameters similarly to what is done in LQR design.
\vspace{5pt}

\begin{theorem}\label{softheorem}
Assume that the linear system (\ref{linear}) is given with $(A,B)$ controllable and $(A,C)$ observable.
Assume further that $Q=Q^T\ge 0, R=R^T>0,$ are such that the pair $(A,\sqrt{Q})$ is observable.
If there is a solution of the LMI
\begin{equation}\label{soflmi0}
\left[
\begin{array}{cc}
Q-PBFC-C^TF^TB^TP+N & P\overline B\\
\overline B^TP & R
\end{array}
\right]>0
\end{equation}
where 
\begin{eqnarray}
N &=& PB(FDR^{-1}+R^{-1}D^TF^T)B^TP,\label{N0}\\
\overline B &=& B(I_m+FD)\label{barb0},
\end{eqnarray}
and $P$ is the unique positive definite solution of the Riccati equation (\ref{Riccati}), then the control law $u=Fy$ is a stabilizing static output feedback for system (\ref{linear}) in the absence of disturbances and the system is quadratically dissipative with supply rate $W(w)=w^TRw$.
Conversely, if the system (\ref{linear}) is static output feedback stabilizable in the absence of disturbances and quadratically dissipative with supply rate $W(w)=w^TRw,$ there always exist $F, P>0, R>0, Q$ verifying LMI (\ref{soflmi0}).
\end{theorem}
%\vspace{10pt}
\proof
Since $R>0$ one can use the Schur complement defined in Lemma \ref{Schurlemma} to rewrite LMI (\ref{soflmi0}) equivalently as
\begin{equation*}
Q-PBFC-C^TF^TB^TP+N-P\overline BR^{-1}\overline B^TP>0
\end{equation*}
Using the definition of $N$ and $\overline B$ from (\ref{N0}) and (\ref{barb0}), respectively, and the Riccati equation (\ref{Riccati}), this inequality can be rewritten as
\begin{equation}\label{preliminaryresult0}
-A^TP-PA-PBFC-C^TF^TB^TP-\overline N >0
\end{equation}
where $\overline N=PBFDR^{-1}D^TF^TB^TP$.
Using Schur complement inequality (\ref{preliminaryresult0}) is equivalent to
\begin{equation}\label{stability0}
\left[
\begin{array}{cc}
(A+BFC)^TP+P(A+BFC) & PBFD\\
D^TF^TB^TP & -R
\end{array}
\right]<0
\end{equation}
We first prove that the LMI (\ref{soflmi0}) or, equivalently, LMI (\ref{stability0}) is a sufficient condition.
Assume that there is a solution of the LMI (\ref{stability0}).
Pre-multiplying this inequality by $v=[x^T~w^T]$ and post-multiplying by $v^T$ yields
\begin{equation}\label{provedissipativity0}
x^TP\dot x+\dot x^TPx -w^TRw <0
\end{equation}
where $\dot x=Ax+Bu$ with $u=FCx+FDw$ are the closed-loop dynamics.
Choosing the storage function $V=x^TPx,$ the inequality (\ref{provedissipativity0}) proves dissipativity with supply rate $W(w)=w^TRw$ according to definition \ref{dissipative}.
In the absence of disturbances one can prove asymptotic stability of the closed-loop system using the Lyapunov function $V=x^TPx$.

We now prove the converse result.
Assume that system (\ref{linear}) is stabilizable by static output feedback in the absence of disturbances and quadratically dissipative with supply rate $W(w)=w^TRw$ for arbitrary $R>0$.
Then there exist matrices $P>0$ and $F$ such that $V=x^TPx$ is a storage Lyapunov function for the closed-loop system with $P, F$ verifying inequality (\ref{stability0}).
Since $R$ is invertible, we can add and subtract the terms $PBR^{-1}B^TP$ and $N$ defined in (\ref{N0}) to the Schur complement of inequality (\ref{stability0}) and then define a matrix $Q$ given by (\ref{Riccati}).
This allows us to rewrite (\ref{stability0}) equivalently as the LMI (\ref{soflmi0}) using Schur complement.
Therefore, there always exist matrices $F, P>0,$ such that for arbitraty $R>0,$ there always exists a $Q=Q(R,P)$ given by (\ref{Riccati}) so that the LMI (\ref{soflmi0}) is satisfied.
\done
\vspace{5pt}
\begin{remark}
When $D=0$ the LMI (\ref{soflmi0}) becomes
\begin{equation}\label{noD}
\left[
\begin{array}{cc}
Q-PBFC-C^TF^TB^TP & PB\\
B^TP & R
\end{array}
\right] >0
\end{equation}
Defining $\bar A=-BFC$ then the constraint (\ref{noD}) becomes the LQR synthesis LMI  (\ref{LMILQR}) for the pair $(\bar A,B)$ \cite{willems71}.\done
\end{remark}
%\vspace{10pt}
\begin{remark}
From the proof of theorem \ref{softheorem}, when $D=0$ and equation (\ref{Riccati}) is satisfied then LMI (\ref{soflmi0}) is equivalently to LMI (\ref{stability0}), i.e,
\begin{equation}\label{stability}
\left[
\begin{array}{cc}
(A+BFC)^TP+P(A+BFC) & 0\\
0 & -R
\end{array}
\right]<0
\end{equation}
which in turn is equivalent to (\ref{LMIforstability}).
Therefore, from the previous remark we can conclude that LMI (\ref{noD}) is equivalent to LMI (\ref{LMIforstability}) when equation (\ref{Riccati}) is satisfied.\done
\end{remark}
%\vspace{10pt}
\begin{remark} 
Removing the positive semi-definite term $N$ from inequality (\ref{soflmi0}) yields a sufficient condition for the design of static output feedback controllers that render the closed-loop system dissipative.\done
\end{remark}
%\vspace{10pt}
\iffalse
\begin{remark}
Please note that $(\alpha I+R)^{-1}=R^{-1}-\alpha R^{-1}(I+\alpha R^{-1})^{-1}R^{-1}$.\done
\end{remark}
\fi
\vspace{5pt}

\begin{corollary}
If $C=I$ and $D=0$ then the LQR gain matrix $F=-R^{-1}B^TP$ is a solution of LMI (\ref{soflmi0}), where $P$ is the unique positive definite solution of Riccati equation (\ref{Riccati}).
%\vspace{10pt}

\proof
Replacing $C=I, D=0,$ and $F=-R^{-1}B^TP$ in LMI (\ref{soflmi}) yields
\begin{equation}\label{statelmi}
\left[
\begin{array}{cc}
Q+PBR^{-1}B^TP+PBR^{-1}B^TP & PB\\
B^TP & R
\end{array}
\right]>0
\end{equation}
Since $R>0$ by Schur complement this inequality is equivalent to
\begin{equation}\label{simplelmi}
Q+PBR^{-1}B^TP>0
\end{equation}
From equation (\ref{Riccati}) we observe that
\begin{equation}
-Q=A^TP+PA-PBR^{-1}B^TP
\end{equation}
so that inequality (\ref{simplelmi}) can be rewritten as
\begin{equation}
(A-BR^{-1}B^TP)^TP+P(A-BR^{-1}B^TP)<0
\end{equation}
For $F=-R^{-1}B^TP$ this inequality is equivalent to
\begin{equation}
(A+BF)^TP+P(A+BF)<0
\end{equation}
which is satisfied because $P$ and $F$ are obtained from the solution of an LQR problem.
\done
\end{corollary}
\vspace{5pt}

\noindent Theorem \ref{softheorem} motivates the following optimization.
\vspace{5pt}

\noindent {\bf Static Output Feedback Optimization 1}
\begin{equation}\label{LMIOpt1}
\begin{array}{rl}
\mbox {\rm max}~{\alpha}~~~~~~~~~~~~~~~~~~~~~~~~~~~~~~~~~~~~~~~~~~~~~~~~~~~~~~~~\\
s.t.  
\left[
\begin{array}{cc}
Q-PBFC-C^TF^TB^TP+N & P\overline B\\
\overline B^TP & R
\end{array}
\right] > \alpha I
\end{array}
\end{equation}
where $I=I_{n+m}$ is the identity matrix of order $n+m$.
\vspace{5pt}
\begin{remark}
If all eigenvalues of $A+BFC$ lie on the left half of the complex plane and $D=0,$ then the closed-loop system is exponentially stable.
From the proof of theorem \ref{softheorem} it is clear that this is guaranteed when $\alpha\ge 0$.
Note however that due to the nature of interior point algorithms all the eigenvalues of $A+BFC$ might lie on the left half of the complex plane even when $\alpha<0$, provided $\alpha$ is close enough to zero.
Therefore, one should always check the eigenvalues of $A+BFC$ after the design of $F$ is performed by the optimization (\ref{LMIOpt1}), even when $\alpha<0$.\done
\end{remark}
%\vspace{10pt}
\begin{remark}
Note that $F$ does not have to be the only variable in LMI (\ref{LMIOpt1}).
It is assumed that the matrix $P>0$ in (\ref{LMIOpt1}) was previously obtained from the solution of an LQR problem using weighting matrices that will be denoted $Q_0$ and $R_0$.
However, although the matrix $P$ must be fixed as the solution of the LQR problem, for fixed $R=R_0$ the matrix $Q$ can also be considered as a variable in problem (\ref{LMIOpt1}), provided one adds the constraint $0\le Q\le Q_0$.
The reason for this is that if the constraint (\ref{LMIOpt1}) is satisfied for $Q\le Q_0$ then it is also satisfied for $Q_0$.
This adds more free variables to the maximization of $\alpha$ than fixing $Q=Q_0$.
Furthermore, it also adds more free variables when compared with the LMI constraint (\ref{LMIforstability}) for the case $D=0$ and is therefore a less conservative condition.
Note that since the LMI  (\ref{LMIOpt1}) is linear in $Q$ convexity is still maintained when considering $Q$ as a variable.\done
\end{remark}

\section{Static Output Feedback with Direct Feedthrough}\label{directfeedthrough}
\noindent Consider now the dynamics of a linear time-invariant system with direct feedthrough term
\begin{eqnarray}\label{linearD}
\dot x &=& Ax + Bu\nonumber\\
y &=& Cx+Du
\end{eqnarray}
where $x\in\RE^n$ is the state, $u\in\RE^m$ is the input, and $y\in\RE^p$ is the output.
Note that this model corresponds to making $w=u$ in system (\ref{linear}).
As before, the pair $(A,B)$ is assumed to be controllable and the pair $(A,C)$ is assumed to be observable.
The static output feedback problem in this section is to find a control input
\begin{equation}\label{sofinput}
u=Fy=FCx+FDu
\end{equation}
which can be rewritten as
\begin{equation}\label{controlaw}
u=(I-FD)^{-1}FCx=\overline Fx
\end{equation}
such that the closed-loop system
\begin{equation}
\dot x = \left[A+B(I-FD)^{-1}FC\right]x
\end{equation}
is asymptotically stable.\footnote{in fact exponentially stable since the system is linear}
\vspace{5pt}

\begin{theorem}\label{softheorem}
Assume that the linear system (\ref{linearD}) is given with $(A,B)$ controllable and $(A,C)$ observable.
Assume further that $Q=Q^T\ge 0, R=R^T>0,$ are such that the pair $(A,\sqrt{Q})$ is observable.
If there is a solution of
\begin{equation}\label{soflmi}
\left[
\begin{array}{ccc}
M & P\overline B & C^TF^T\\
\overline B^TP & R & 0_{m\times p}\\
FC & 0_{p\times m} & \overline R
\end{array}
\right]>0
\end{equation}
where 
\begin{eqnarray}
M &=& Q-PBFC-C^TF^TB^TP+N\label{M}\\
N &=& PB(FDR^{-1}+R^{-1}D^TF^T)B^TP,\label{N}\\
\overline B &=& B(I+FD)\label{barb},\\
\overline R &=& (I-FD)R^{-1}(I-D^TF^T)\label{Rbar}
\end{eqnarray}
and $P$ is the unique positive definite solution of the Riccati equation (\ref{Riccati}), then the control law (\ref{controlaw}) is a stabilizing static output feedback for system (\ref{linearD}).
Conversely, if the system (\ref{linearD}) is static output feedback stabilizable, there always exist $F, P>0, R>0, Q$ verifying (\ref{soflmi}).
\end{theorem}
%\vspace{10pt}
\proof
Since $R>0$ one can use the Schur complement defined in Lemma \ref{Schurlemma} to rewrite (\ref{soflmi}) as
\begin{equation}\label{firststep}
M-P\overline BR^{-1}\overline B^TP-C^TF^T\overline R^{-1}FC>0
\end{equation}
Using the definition of $M, N$ and $\overline B$ from (\ref{M}), (\ref{N}) and (\ref{barb}), respectively, and the Riccati equation (\ref{Riccati}), this inequality can be equivalently written as
\begin{equation}\label{preliminaryresult}
(A+BFC)^TP+P(A+BFC)+\overline N+C^TF^T\overline R^{-1}FC <0
\end{equation}
where $\overline N=PBFDR^{-1}D^TF^TB^TP$.
Defining $A_{cl}=A+BFC$ and using the expression (\ref{Rbar}) for $\overline R$ and the Schur complement, inequality (\ref{preliminaryresult}) is equivalent to
\begin{equation}\label{stability}
\left[
\begin{array}{cc}
A_{cl}^TP+PA_{cl}+\overline F^TR\overline F & PBFD\\
D^TF^TB^TP & -R
\end{array}
\right]<0
\end{equation}
where $\overline F$ was defined in (\ref{controlaw}).
Pre-multiplying (\ref{stability}) by $v=[x^T~u^T]$ and post-multiplying by $v^T$ yields
\begin{equation}\label{provedissipativity}
x^TP\dot x+\dot x^TPx +x^T\overline F^TR\overline Fx-u^TRu <0
\end{equation}
where $\dot x=Ax+Bu$ with $u$ given by (\ref{controlaw}) are the closed-loop dynamics.
We have therefore shown that inequality (\ref{soflmi}) is equivalent to inequality (\ref{provedissipativity}) provided that the Riccati equation (\ref{Riccati}) is satisfied.
Since the last two terms cancel out, inequality (\ref{provedissipativity}) becomes
\begin{equation}\label{stabilityproof}
x^TP\dot x+\dot x^TPx <0
\end{equation}

We now prove that the inequality (\ref{soflmi}) is a sufficient condition.
Under the stated assumptions there is a unique solution $P>0$ of the Riccati equation (\ref{Riccati}).
Assume that there is also solution of the inequality (\ref{soflmi}).
Choosing the Lyapunov function $V=x^TPx,$ the inequality (\ref{provedissipativity}) proves asymptotic stability of the closed-loop system.
To prove the converse result assume that system (\ref{linearD}) is stabilizable by static output feedback.
Then there exist matrices $P>0$ and $F$ such that $V=x^TPx$ is a Lyapunov function for the closed-loop system with $P, F$ verifying inequality (\ref{stabilityproof}).
Adding and subtracting the term $u^TRu$ with $u=\overline Fx$ yields inequality (\ref{provedissipativity}).
From the fact that $u=FCx+FDu$, defining $v=[x^T~u^T]$ this inequality can be rewritten as (\ref{stability}) because left multiplying (\ref{stability}) by $v^T$ and right multiplying it by $v$ yields (\ref{provedissipativity}).
Since $R$ is invertible, we can add and subtract the terms $PBR^{-1}B^TP$ and $N$ defined in (\ref{N}) to the Schur complement of inequality (\ref{stability}) and then define a matrix $Q$ given by (\ref{Riccati}).
This allows us to rewrite (\ref{stability}) as the LMI (\ref{firststep}).
Using Schur coomplement one can equivalently write (\ref{firststep}) as (\ref{soflmi}).
Therefore, there always exist matrices $F, P>0,$ such that for arbitraty $R>0,$ there always exists a $Q=Q(R,P)$ given by (\ref{Riccati}) so that the LMI (\ref{soflmi}) is satisfied.
\done
\vspace{5pt}

\noindent Since the inequality (\ref{soflmi}) is a Bilinear Mattrix Inequality (BMI) the following corollary presents a sufficient condition for static output feedback as an LMI.
\vspace{5pt}

\begin{corollary}
Under the same assumptions of theorem \ref{softheorem} if there is a solution of inequality (\ref{soflmi}) with $\overline R$ replaced by $I-FDR^{-1}-R^{-1}D^TF^T$ and $P$ is the unique positive definite solution of the Riccati equation (\ref{Riccati}), then the control law (\ref{controlaw}) is a stabilizing static output feedback for system (\ref{linearD}).
\end{corollary}
%\vspace{10pt}
\proof
The result follows because the term that was removed from $\overline R$ is the matrix $FDR^{-1}D^TF^T$, which is positive semidefinite.
\done
\vspace{5pt}
\begin{remark} 
Notice that from the proof of theorem \ref{softheorem} the inequality (\ref{soflmi}) is equivalent to inequality (\ref{stability}), provided that the Riccati equation (\ref{Riccati}) is satisfied. Furthermore, for $D=0$ if inequality (\ref{stability}) is satisfied it implies that inequality (\ref{stability0}) is also satisfied.\done
\end{remark}
\section{Examples}\label{examples}
The software CVX \cite{cvx} was used to solve the optimization problem (\ref{LMIOpt1}) in all examples.
\iffalse
Many examples taken from several different references available in the literature are solved successfully in this section.
The software CVX \cite{cvx} was used to solve the optimization problem (\ref{LMIOpt1}).
% and (\ref{LMIOpt2}).
\fi
\iffalse
\begin{example}\cite{cao98,madeira16}
For the system with matrices
\begin{eqnarray*}
A &=&\left[
\begin{array}{cc}
0 & 1\\
1 & 0
\end{array}
\right], 
B = \left[
\begin{array}{c}
1\\
0
\end{array}
\right],
C = \left[
\begin{array}{c}
1\\
10
\end{array}
\right]^T
\end{eqnarray*}
using $Q=100I_2, R=1,$ (\ref{LMIOpt1}) yields $F=-1.2287$.
\end{example}
%\vspace{2.5pt}
\begin{example}\cite{mesbahi99}
For the system with matrices
\begin{eqnarray*}
A &=&\left[
\begin{array}{cc}
1 & 1.005\\
-1.005 & 0
\end{array}
\right], 
B = \left[
\begin{array}{c}
0\\
1
\end{array}
\right],
C = \left[
\begin{array}{c}
0\\
1
\end{array}
\right]^T
\end{eqnarray*}
using $Q=\operatorname{diag}([10^{-8}~10^{-11}]), R=1.5\times 10^{-8},$ the solution of (\ref{LMIOpt1}) is $F=-1.0007$.
\end{example}
%\vspace{2.5pt}
\fi
\begin{example}\cite{anderson75}
For the system with matrices
\begin{eqnarray*}
A =\left[
\begin{array}{ccc}
0 & 1 & 0\\
0 & 1 & 1\\
0 & 13 & 0
\end{array}
\right], 
B = \left[
\begin{array}{c}
0\\
0\\
1
\end{array}
\right] &,&\\
C = \left[
\begin{array}{ccc}
0 & 5 & -1\\
-1 & -1 & 0
\end{array}
\right] &,&
\end{eqnarray*}
using $Q=\operatorname{diag}[1~3~0.1], R=10^{-4},$ the solution of (\ref{LMIOpt1}) is $F=[6.8981~84.9224]$.
\end{example}
%\vspace{2.5pt}
\iffalse
\begin{example}\cite{madeira16}
For the system with matrices
\begin{eqnarray*}
A =\left[
\begin{array}{ccc}
-4 & 1 & 0\\
7 & 0 & 0\\
1 & 0 & 0
\end{array}
\right], 
B = \left[
\begin{array}{c}
-0.1\\
0\\
10
\end{array}
\right] &,&\\
C = \left[
\begin{array}{ccc}
1 & 0 & 1\\
0 & 0 & 1
\end{array}
\right] &,&
\end{eqnarray*}
using $Q=100I_3, R=1,$ the solution of (\ref{LMIOpt1}) is $F=[-4.1787~-5.6663]$.
\end{example}
%\vspace{2.5pt}
\fi
\begin{example}\cite{geromel94,cao98,mesbahi99}
For the triplet
\begin{eqnarray*}
A =\left[
\begin{array}{cccc}
-0.0366 & 0.0271 & 0.0188 & -0.4555\\
0.0482 & -1.0100 & 0.0024 & -4.0208\\
0.1002 & 0.3681 & -0.7070 & 1.4200\\
0 & 0 & 1 & 0
\end{array}
\right]&,&\\ 
B = \left[
\begin{array}{cc}
0.4422 & 0.1761\\
3.5446 & -7.5922\\
-5.5200 & 4.4900\\
0 & 0
\end{array}
\right],
C = \left[
\begin{array}{cccc}
0 & 1 & 0 & 0
\end{array}
\right] &,&
\end{eqnarray*}
using $Q=I_4, R=I_2,$ (\ref{LMIOpt1}) yields $F=[2.8334~8.8618]$.
\end{example}
%\vspace{2.5pt}
\iffalse
\begin{example}\cite{karimi13}
Consider the linear system
\begin{eqnarray*}
A =\left[
\begin{array}{ccccc}
-4 & 0 & -2 & 0 & 0\\
0 & -2 & 0 & 2 & 0\\
0 & 0 & -2 & 0 & -1\\
0 & -2 & 0 & -1 & 0\\
3 & 0 & -2 & 0 & -1
\end{array}
\right]&,&\\ 
B = \left[
\begin{array}{ccc}
1 & 0 & 0\\
1 & 0 & 0\\
0 & 0 & 0\\
0 & 1 & 0\\
0 & 0 & 1
\end{array}
\right],
C^T = \left[
\begin{array}{ccc}
1 & 0 & 0\\
0 & 1 & 0\\
0 & 0 & 0\\
0 & 0 & 0\\
0 & 0 & 1
\end{array}
\right] &,&\\
B_w=\left[
\begin{array}{ccccc}
1 & 1 & 1 & 1 & 1
\end{array}
\right]^T &,&\\
C_z=\left[
\begin{array}{c}
I_5\\
0_{3\times 5}
\end{array}
\right],\quad
D_z=\left[
\begin{array}{c}
0_{5\times 3}\\
I_3
\end{array}
\right] &,&
\end{eqnarray*}
using $Q=\operatorname{diag}[2.5~0.95~10~6.9~1], R=1.4I_3,$ the solution of (\ref{LMIOpt2}) is
\begin{eqnarray*}
F =\left[
\begin{array}{ccc}
-0.4883 &  -0.3979 &  -0.4954\\
 -0.2727 &   0.2914 &  -0.2113\\
-0.4481 &   0.2337 &  -1.1238
\end{array}
\right], 
\end{eqnarray*}
and the ${\mathcal L}_2$ gain is guaranteed to verify $\bar\gamma\le 0.9641$.
The solution from \cite{karimi13} yields an ${\mathcal L}_2$ gain of $1.2084$.
Note that our bound on the ${\mathcal L}_2$ gain is $20\%$ smaller and it guarantees attenuation of disturbances since $\bar\gamma<1$.
\end{example}
\fi
\bibliographystyle{ieeetr}
\bibliography{references}
\end{document}